# Signatures of disorder in the minimum conductivity of graphene

*Yang Sui[1,2]\*, Tony Low[1,3], Mark Lundstrom[1,3], and Joerg Appenzeller[1,2]*

[1]School of Electrical and Computer Engineering, [2]Birck Nanotechnology Center, [3]Network for Computational Nanoelectronics, Purdue University, West Lafayette, Indiana 47907, USA

suiyang2000@gmail.com

ABSTRACT: Graphene has been proposed as a promising material for future nanoelectronics because of its unique electronic properties. Understanding the scaling behavior of this new nanomaterial under common experimental conditions is of critical importance for developing graphene-based nanoscale devices. We present a comprehensive experimental and theoretical study on the influence of edge disorder and bulk disorder on the minimum conductivity of graphene ribbons. For the first time, we discovered a strong non-monotonic size scaling behavior featuring a peak and saturation minimum conductivity. Through extensive numerical simulations and analysis, we are able to attribute these features to the amount of edge and bulk disorder in graphene devices. This study elucidates the quantum transport mechanisms in realistic experimental graphene systems, which can be used as a guideline for designing graphene-based nanoscale devices with improved performance.

Conductivity of a material offers insights into the energy band structure and exhibits different behaviors depending on the scattering mechanisms involved. Conductivity is independent of geometry for conventional bulk materials, but scales with length for low-dimensional materials under ballistic transport conditions (e.g. carbon nanotubes)[1], where no scattering occurs within the channel. The situation becomes more intriguing for graphene at the Dirac point[2-4], where the density of states (DOS) is zero. Despite zero carrier density, the conduction takes place by the evanescent modes tunnelling through the Dirac point and leads to a counterintuitive finite minimum conductivity ($\sigma_{min}$)[5,6]. In the ideal disorder-free case, $\sigma_{min}$ scales monotonically with the width-to-length ratio ($W/L$) of the channel[6]. In reality, disorder in experimental graphene structures has been found to be inevitable and profoundly impacts carrier transport[7-9], particularly for the off-state. However, there has not been a comprehensive understanding on the scaling of $\sigma_{min}$ over a broad width and length parameter space, where disorder dictates the transport behavior. Here we report an experimental investigation of graphene ribbons in which $\sigma_{min}$ is governed mainly by the impacts of edge and bulk disorder. In distinct contrast to the disorder-free case[6,10], our investigation reveals a strong *non-monotonic W/L* dependence along with an overall enhancement in $\sigma_{min}$, characterized by a *peak* and *saturation*. Our model for the disordered graphene system agrees quantitatively with the experiments. In addition, our framework accommodates previous findings[10] on minimum conductivity for graphene ribbons in different geometrical limits. This study probes into the behavior of disordered graphene system, elucidates the transport mechanism, and provides insights for improving the performance of graphene nanoelectronics.

Over the years, experiments have provided evidence that disorder significantly impacts the off-state carrier transport in graphene structures[9]. Minimum conductivity of graphene, initially thought to be a universal constant[3,11,12], is only so when external perturbations such as disorder and most notably contacts are absent[10,13-15]. However, perturbations are ubiquitous for experimental graphene devices, and their impact on the finite size scaling of $\sigma_{min}$ is unclear due to the lack of experimental data in the disorder dominated regime. To address this issue, we have fabricated bottom-gated graphene devices with dimensions larger than typical length scales of disorder[8] and covered a wide range of width-to-length ratio ($W/L$ = 0.1-2.6). A two-probe setup rather than four-probe is employed since only the former explicitly accounts for the contacts, which are the origins of the evanescent modes and responsible for the observed width scaling behavior[5,6].

Throughout this entire work, we focus on single-layer graphene devices on $SiO_2$/Si substrates. Figure 1a shows a Scanning Electron Microscopy (SEM) micrograph of the structure under investigation. Typical transfer characteristics at different temperatures are shown in Figure 1b, where $\sigma_{min}$ shows little temperature dependence. The width-to-length ratio is used as a metric to investigate the scaling behavior of $\sigma_{min}$ (discussed in Supplement I), with the experimental findings shown in Figure 1c. Both the transfer length ($L_T$) and the contact resistance ($R_C$) are considered when evaluating $L$ and $\sigma_{min}$ (Supplement II). The following distinct features are immediately noticed: (i) $\sigma_{min}$ has a *non-monotonic* dependence on $W/L$, with $\sigma_{peak} \sim 8$ ($4e^2/\pi h$) at $W/L \approx 0.5$; (ii) $\sigma_{min}$ saturates for large $W/L$, denoted as $\sigma_{sat} \sim 4$ ($4e^2/\pi h$). Very few previous reports can be found on the geometrical scaling of $\sigma_{min}$. Figure 1c includes previous published data[10] "+" and "Δ" that exhibit an apparent monotonic scaling behavior, just as predicted for disorder-free graphene[6]. By studying devices much larger than typical dimensions of disorder and probing into a much broader $W/L$ regime, we have uncovered new features in the scaling behavior of $\sigma_{min}$, characterized by $\sigma_{peak}$ and $\sigma_{sat}$.

First, it is useful to recall that there are two types of states in a graphene ribbon, namely edge states and bulk states. Edge states are highly conductive propagating states residing along non-armchair edges[16-18]. The conductance from edge states is independent of $W$, and therefore would dominate as $W/L$ approaches zero and lead to a diverging $\sigma_{min}$[6]. Bulk states on the other hand are less conductive evanescent states with the number of states proportional to $W$. In the large $W/L$ regime, bulk states dominate the transport and lead to a constant $\sigma_{min}$ of $4e^2/\pi h$ in the disorder-free case[5,6]. Combining these two types of states is the key to understand the experimental observations. Note that both types of states are involved in transport through zigzag ribbons, whereas only bulk states are responsible for transport in armchair ribbons.

Edge disorder is a result of the inevitable edge roughness from imperfect cleaving or top-down fabrication. It is known to cause gap-like transport behavior due to localization or quantum dot formation[19-22]. A zigzag edge is more common for graphene ribbons than an armchair edge since ribbons with edges "between" zigzag and armchair tend to acquire zigzag-type boundaries[16]. In the absence of disorder, $\sigma_{min}$ is predicted to increase or decrease monotonically with $W/L$ for armchair or zigzag ribbons[6] (black curves in Figure 2a). We have performed quantum transport simulations using a Non-Equilibrium Green's Function (NEGF) formalism to model the edge and bulk disorder in 2D graphene sheets, with the numerical implementations described in Supplement IV and V. All simulations are performed for T = 0 K (see supplement III). Indeed, edge disorder immediately gives rise to a non-monotonic behavior for zigzag ribbons, while $\sigma_{min}$ for armchair ribbons remains monotonic as a function of $W/L$, albeit with a lower conductivity (Figure 2a). Interestingly, our experimental devices (Figure 1c) exhibit a scaling behavior that is consistent only with edge-disordered zigzag ribbons, and none of the devices fall into the "armchair" trend as predicted by the green circles in Figure 2a. This observation can be viewed as a manifestation of the fact that the zigzag boundary condition, where the edge is characterized by a majority sub-lattice and hence accommodates edge states, is generic for experimental graphene ribbons.

An analytical description for the impact of edge disorder can be obtained by considering a width-dependent transport gap. $\sigma_{min}$ for a zigzag ribbon with edge roughness can be described as follows[6]:

$$\sigma_{min} = \frac{\pi L}{W}\left\{\sum_{n=2}^{\infty}\cosh^{-2}\left(\frac{n\pi}{W}L\right) + \frac{1}{2}\cosh^{-2}\left(\frac{\varepsilon_{loc}}{\hbar v_f}L\right)\right\} \quad (1)$$

given in units of $4e^2/\pi h$. $\varepsilon_{loc} = \alpha/W^\beta$ is the Anderson localization-induced transport gap with $\alpha \sim 1$ ÅeV and $\beta \sim 1$, as estimated in[20]. Although temperature should have an effect on $\varepsilon_{loc}$ due to dephasing, however, the thermal broadening due to Fermi function smearing is negligible as discussed in Supplement III. Since $\varepsilon_{loc}$ increases with decreasing width, $\sigma_{min}$ for narrow ribbons is more strongly suppressed than for wide ribbons, resulting in decreasing $\sigma_{min}$ values for small $W/L$. Furthermore, the edge states wavefunction decays exponentially from the edges with a characteristic length proportional to $W$ at a given Fermi energy ($E_F$)[23], making the narrower ribbons more susceptible to edge disorder. Another way to form a transport gap in graphene nanoribbons (GNRs) is to combine small confinement-induced energy gaps with potential inhomogeneities[19,21,22]. We believe that the transport gap in our case is mainly due to edge localization for the following reasons: First, the smallest $W$ (~ 100 nm) is much larger than the effective diameter of a charging island (~ 20 nm) near the Dirac point[8], and is unlikely to form isolated quantum dots. Secondly, the confinement-induced energy gap for $W \geq 100$ nm is too small to block the band-to-band tunnelling current. Therefore, it is likely that edge roughness scattering is responsible for $\sigma_{min}$ reduction for zigzag ribbons of small $W/L$. The length scaling of $\sigma_{peak}$ is found to follow a power-law relationship, as shown in Figure 2b. Extrapolation to $L = 1$ μm yields $\sigma_{peak} \approx 4.2$ ($4e^2/\pi h$) for reasonable assumptions of root-mean-square edge roughness ($RMS$ = 7.5 nm) and autocorrelation length ($AL$ = 2 nm)[24,25]. Note that since the peak position (not the peak height) is roughly a constant when the $RMS$ roughness is varied, we deduce that $\sigma_{peak} \propto W^{0.46}$ (as $W/L \approx$ constant at the peaks) – a finding that will become relevant in subsequent discussions.

Bulk disorder in graphene arises from charge inhomogeneity at the graphene-substrate interface, with a typical potential fluctuation of ± 40 meV[8], the origin of which could be due to charge impurities on $SiO_2$. Local electron- and hole-rich regions, called electron-hole puddles, are induced in graphene with typical diameters of ~150 nm[8]. For any given $E_F$ line-up in the off-state, percolation paths exist from the source to the drain through connected electron and hole puddles. As a result, current flows preferably through the regions of unperturbed potential to avoid the additional resistance associated with the pn interfaces at puddle boundaries[26], as demonstrated by Figure 3a. Therefore, bulk disorder has a strong influence on $\sigma_{min}$.

Figure 3b shows the NEGF simulation results for $\sigma_{min}$ vs. $W/L$ for ribbons with bulk disorder. Adding bulk disorder to graphene leads to higher conductivity[27], which is exactly the opposite trend compared to other material systems. The reason for this unique behavior lies in the fact that graphene is a gapless material, where carriers can tunnel almost unimpeded from the conduction band into the valence band.

Different $E_F$ line-ups with respect to the Dirac point lead to a variation of DOS and thus conductivity in different puddles. The average DOS in the puddles increases as the magnitude of bulk disorder increases due to the linear energy dispersion of graphene. This leads to an enhancement in $\sigma_{min}$ with increasing bulk disorder over the entire $W/L$ range (Figure 3b). Effective medium theory is able to capture the main features of this phenomenon (Figure 3c), with equations in Supplement VI. It is a remarkable coincidence that the saturation feature of the bulk states conductivity persists for both disorder-free and bulk-disordered ribbons, despite different mechanisms involved. The former is due to evanescent transport, while the latter is dominated by percolation. For smaller devices (e.g. $L < 500$ nm) that only contain a few puddles, the potential is relatively uniform across the channel and the impact of bulk disorder is minimal. As a result, $\sigma_{min}$ is markedly lower than for large devices[10], just like the disorder-free case[6] (e.g. "+" in Figure 1c). Using the effective medium model, we obtain an effective conductivity $\sigma_{eff} \propto W^{0.16}$ for $W/L < 1$, assuming $\sigma_1/\sigma_2 = 0.01$ and $L \approx 30 \, \mu_R$ (puddle radius). Rigorous NEGF simulations assuming reasonable bulk disorder parameters extrapolate to $\sigma_{sat} \approx 4.3$ ($4e^2/\pi h$) for $L = 1$ μm. Quantitative agreement between experiments and NEGF simulations for the peak and saturation $\sigma_{min}$ is reached, with details in Supplement VII.

Finally, we discuss the implications of this work for nanoscale graphene devices. It is a major hurdle to realize high-performance graphene transistors due to the absence of a bandgap. Our study provides valuable insights in this context. Figure 4a shows the width scaling of the on- and off-state conductance ($G_{on}$ and $G_{off}$). Interestingly, while $G_{on} \propto W$, $G_{off}$ follows an unconventional trend $G_{off} \propto W^{1.3}$. We attribute this stronger suppression in $G_{off}$ as $W$ decreases to the following mechanisms: (i) the number of rough edges per unit width is increased, thereby suppressing the highly conductive edge states; (ii) some percolation paths are terminated. The particular $G_{off}-W$ dependence can be rationalized knowing $\sigma_{min} \propto W^{0.16-0.46}$ for $W/L < 0.5$ and $G_{off} = \sigma_{min} \cdot W/L \propto W^{1.16-1.46}$. Width reduction without compromising the on-current can be accomplished by cutting nanometer-scale incisions along the transport direction (Figure 4b). We have observed at least a two- to three-fold improvement in on/off ratio by moderately reducing the effective width from 1 μm to 100 nm, as demonstrated in Figure 4a. This also allows us to scale the channel width without modifying the on/off ratio to facilitate circuit level design requirements. In a similar fashion, the recently demonstrated graphene nanomesh works under the same principle, and was found to achieve an on/off ratio of ~160[28]. Exploiting these geometries will allow designing better graphene-based nanoelectronics such as RF devices and bio-sensors[29,30].

NOTE: Our study should have relevance to the recently demonstrated atomically flat graphene on boron nitride system[31]. In fact, we expect a stronger non-monotonic dependence of $\sigma_{min}$ as a function of $W$ due to suppression of bulk disorder in such a system.

METHODS: Graphene flakes are mechanically exfoliated onto 90 nm $SiO_2$/Si substrates from Highly Oriented Pyrolytic Graphite (HOPG, NT-MDT). Source and drain contacts are patterned by e-beam lithography, followed by e-beam evaporation and lift-off of a stack of Ti/Pd/Au (10 nm/30 nm/20 nm) metals. In order to prevent damages to the graphene ribbons from the e-beam patterning process, we choose graphene flakes of various widths and lengths as naturally cleaved from the exfoliation process. In this way, there are no further modifications on the graphene ribbons widths. Devices with $W$ = 0.1-4.9 μm and physical lengths ($L_{SEM}$) of 1.0-1.5 μm are created from single-layer graphene flakes. The thickness of the graphene flakes is determined by optical analysis and Atomic Force Microscopy (AFM). Electrical measurements of the devices are carried out at T = 4-300 K under vacuum (1x10$^{-7}$ torr). The field-effect mobility of typical devices is ~ 4000 cm$^2$/Vs as determined from room temperature transfer characteristics.

ACKNOWLEDGMENT: The authors thank S. Datta, M. Alam, Z. Jacob, and F. Guinea for discussions, and J. T. Smith for assistance with the graphics. The work was supported by INDEX (NRI) and Intel Corp., with computational resources from NCN.

REFERENCES:

1. Javey, A.; Guo, J.; Wang, Q.; Lundstrom, M.; Dai, H. J. *Nature* **2003**, 424, 654.

2. Novoselov, K. S.; Geim, A. K.; Morozov, S. V.; Jiang, D.; Zhang, Y.; Dubonos, S. V.; Grigorieva, I. V.; Firsov, A. A. *Science* **2004**, 306, 666.

3. Novoselov, K. S.; Geim, A. K.; Morozov, S. V.; Jiang, D.; Katsnelson, M. I.; Grigorieva, I. V.; Dubonos, S. V.; Firsov, A. A. *Nature* **2005**, 438, 197.

4. Zhang, Y. B.; Tan, Y. W.; Stormer, H. L.; Kim, P. *Nature* **2005**, 438, 201.

5. Katsnelson, M. I. *European Physical Journal B* **2006**, 51, 157.

6. Tworzydlo, J.; Trauzettel, B.; Titov, M.; Rycerz, A.; Beenakker, C. W. J. *Physical Review Letters* **2006**, 96, 246802.

7. Sui, Y.; Appenzeller, J. *Nano Letters* **2009**, 9, 2973.

8. Martin, J.; Akerman, N.; Ulbricht, G.; Lohmann, T.; Smet, J. H.; Von Klitzing, K.; Yacoby, A. *Nature Physics* **2008**, 4, 144.

9. Neto, A. H. C.; Guinea, F.; Peres, N. M. R.; Novoselov, K. S.; Geim, A. K. *Reviews of Modern Physics* **2009**, 81, 109.

10. Miao, F.; Wijeratne, S.; Zhang, Y.; Coskun, U. C.; Bao, W.; Lau, C. N. *Science* **2007**, 317, 1530.

11. Khveshchenko, D. V. *Physical Review Letters* **2006**, 97, 036802.

12. Nomura, K.; MacDonald, A. H. *Physical Review Letters* **2007**, 98, 076602.

13. Tan, Y. W.; Zhang, Y.; Bolotin, K.; Zhao, Y.; Adam, S.; Hwang, E. H.; Das Sarma, S.; Stormer, H. L.; Kim, P. *Physical Review Letters* **2007**, 99, 246803.

14. Geim, A. K.; Novoselov, K. S. *Nature Materials* **2007**, 6, 183.

15. Chen, J. H.; Jang, C.; Adam, S.; Fuhrer, M. S.; Williams, E. D.; Ishigami, M. *Nature Physics* **2008**, 4, 377.

16. Akhmerov, A. R.; Beenakker, C. W. J. *Physical Review B* **2008**, 77, 085423.

17. Kobayashi, Y.; Fukui, K.; Enoki, T.; Kusakabe, K.; Kaburagi, Y. *Physical Review B* **2005**, 71, 193406.

18. Fujita, M.; Wakabayashi, K.; Nakada, K.; Kusakabe, K. ; *Journal of Physical Society Japan* **1996**, 65, 1920.

19. Gallagher, P., Todd, K.; Goldhaber-Gordon, D. *Physical Review B* **2010**, 81, 115409.

20. Mucciolo, E. R.; Neto, A. H. C.; Lewenkopf, C. H. *Physical Review B* **2009**, 79, 075407.

21. Sols, F.; Guinea, F.; Neto, A. H. C. *Physical Review Letters* **2007**, 99, 166803.

22. Han, M. Y.; Brant, J. C.; Kim, P. *Physical Review Letters* **2010**, 104, 056801.

23. Brey, L.; Fertig, H. A. *Physical Review B* **2006**, 73, 235411.

24. Gupta, A. K.; Russin, T. J.; Gutierrez, H. R.; Eklund, P. C. *Acs Nano* **2009**, 3, 45.

25. Tapaszto, L.; Dobrik, G.; Lambin, P.; Biro, L. P. *Nature Nanotechnology* **2008**, 3, 397.

26. Cheianov, V. V.;  Fal'ko, V. *Physical Review B* **2006**, 74, 041403.

27. Bardarson, J. H.; Tworzydlo, J.; Brouwer, P. W.; Beenakker, C. W. J. *Physical Review Letters* **2007**, 99, 106801.

28. Bai, J.; Zhong, X.; Jiang, S.; Huang, Y.; Duan, X. *Nature Nanotechnology* **2010**, 5, 190.

29. Lin, Y. M.; Dimitrakopoulos, C.; Jenkins, K. A.; Farmer, D. B.; Chiu, H. Y.; Grill, A.; Avouris, P. *Science* **2010**, 327, 662.

30. Schedin, F.; Geim, A. K.; Morozov, S. V.; Hill, E. W.; Blake, P.; Katsnelson, M. I.; Novoselov, K. *Nature Materials* **2007**, 6, 652.

31. Dean, C. R.; Young, A. F.; Meric, I.; Lee, C.; Wang, L.; Sorgenfrei, S.; Watanabe, K.; Taniguchi, T.; Kim, P.; Shepard, K. L.; Hone, J. *Nature Nanotechnology* **2010**, 5, 722.

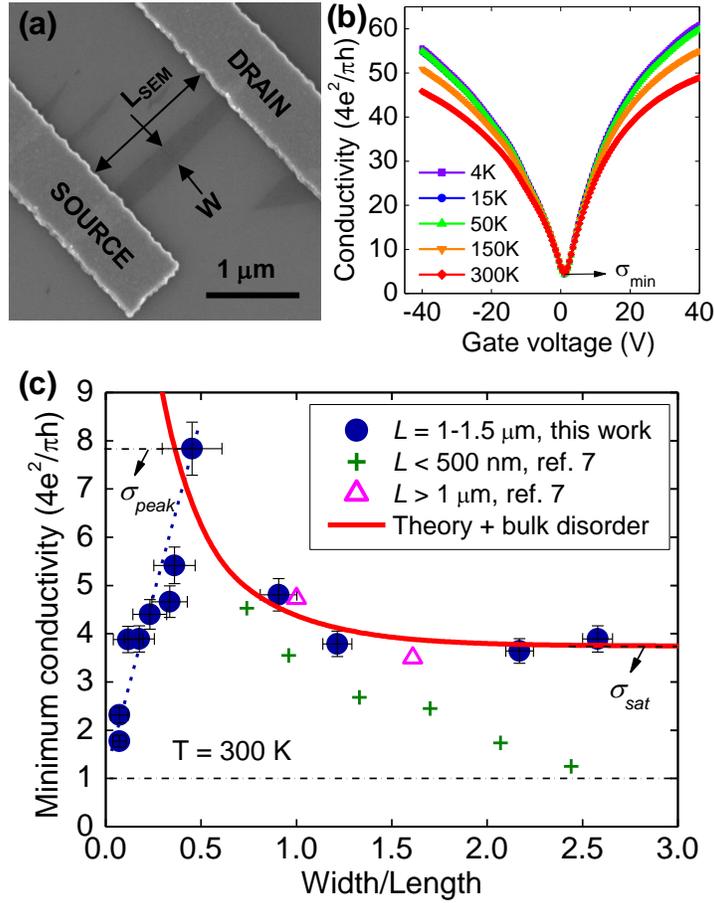

**Figure 1.** (a) SEM micrograph of a typical graphene device. (b) Transfer characteristics of a graphene device at T = 4-300 K. (c) Experimental observations of the geometrical scaling behavior of disordered graphene ribbons. The blue circles are our data from ribbons of various geometries, $W$ = 0.1-4.9 μm and $L$ = 1-1.5 μm. The green "+" and purple "Δ" are reproductions from a previous report[10]. The red solid curve indicates theory for disorder-free graphene ribbons[6] with the enhancement from bulk disorder.

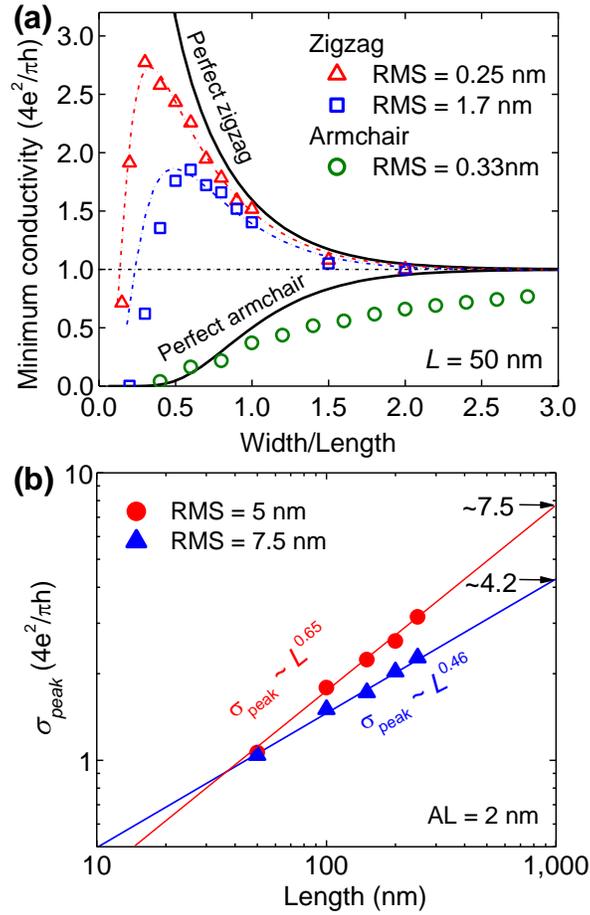

**Figure 2.** (a) Modeling results for $\sigma_{min}$ in edge-disordered graphene ribbons. Symbols are from NEGF simulations for zigzag and armchair ribbons with different *RMS*. *AL* =1 nm. Dash-dotted lines are obtained analytically from eq. (1) with α as a fitting parameter. The solid black curves are obtained from theory for disorder-free ribbons[6]. (b) Length scaling simulation for $\sigma_{peak}$. $W/L \approx 0.5$ and *AL* = 2 nm. Power-law extrapolation to $L = 1$ μm yields $\sigma_{peak} = 7.5$ and 4.2 ($4e^2/\pi h$) for *RMS* = 5 and 7.5 nm, respectively. Note: The motivation for the length scaling study is that it is impractical to simulate live size devices (several μm) due to the limitation of computing power.

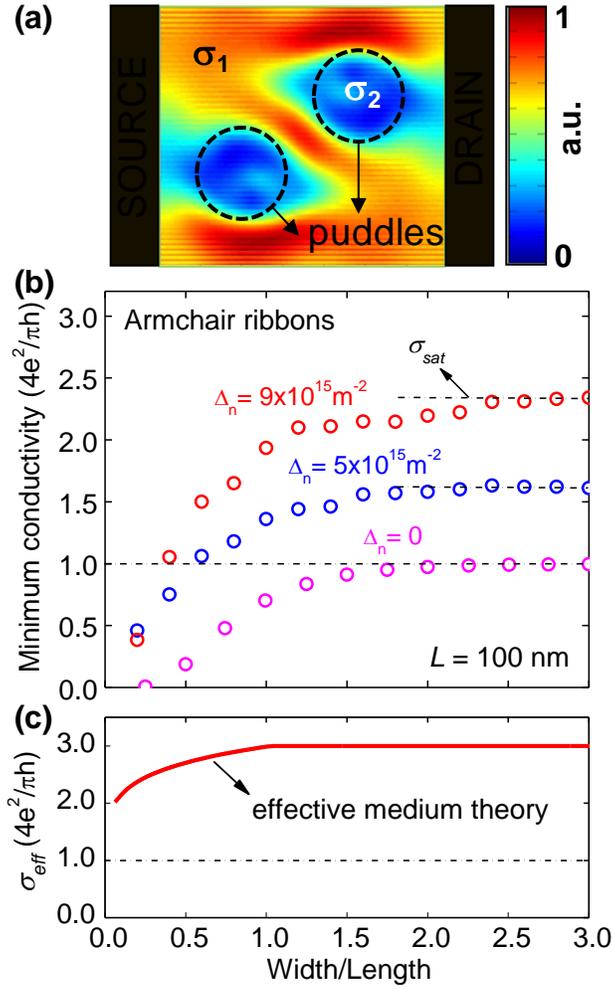

**Figure 3.** (a) Current density in a graphene channel with two puddles (marked by circles). $\sigma_1$ and $\sigma_2$ indicate regions of high and low conductivity. (b) Modeling results for $\sigma_{min}$ in bulk-disordered graphene ribbons. Symbols are from NEGF simulations for different charge density fluctuations ($\Delta_n$). Armchair ribbons are used, but the results can be extended to all types of ribbons. $L$ = 100 nm, $\mu_R$ = 3 nm, $RMS$ ~ 0.25 nm, and $AL$ ~ 1 nm. (c) $\sigma_{eff}$ vs. $W/L$ according to the effective medium model (equations in Supplement VI).

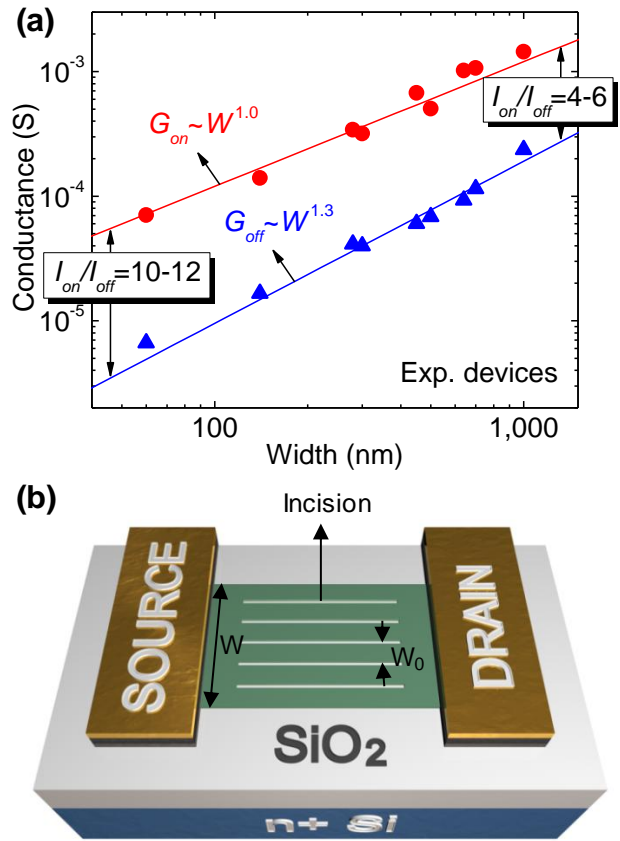

**Figure 4.** (a) Experimental observations of on-state and off-state conductance ($G$) – width scaling. $L \sim 2$ μm and $W/L \leq 0.5$ for all devices, belonging to the ascending trend in Figure 1c. (b) Structure of a proposed graphene transistor with mesoscopic incisions for improved on/off ratio and scalability.

# Supplementary Information

*Yang Sui[1,2*], Tony Low[1,3], Mark Lundstrom[1,3], and Joerg Appenzeller[1,2]*

[1]School of Electrical and Computer Engineering, [2]Birck Nanotechnology Center, [3]Network for Computational Nanoelectronics, Purdue University, West Lafayette, Indiana 47907, USA

**I. Discussion for employing the width-to-length ratio as the scaling metric**

The width-to-length ratio has been frequently used to represent graphene ribbon geometry for studying $\sigma_{min}$[1-3]. In the disordered regime, $\sigma_{min}$ is not governed by $W/L$ but rather by $W$ and $L$. However, it is impractical to cover the whole parameter space of all widths and lengths. Small devices ($L < 500$ nm) can be viewed as disorder-free graphene since the characteristic lengths of disorder (e.g. diameter of electron-hole puddle) are comparable to the device size, resulting in weak impact from the disorder. We choose to investigate in larger devices ($L \geq 1$ μm), where disorder dominates the off-state transport and reveals its signatures. Our devices lengths are rather constant ($L = 1\text{-}1.5$ μm) and widths vary largely from 100 nm to 4.9 μm, covering $W/L = 0.1\text{-}2.6$. Devices of this dimension are most relevant to nanoelectronics. We employ $W/L$ as the scaling metric in order to compare with previous theoretical studies[2], as well as with experimental reports on devices of different sizes[1]. Although $W/L$ is used to represent the device geometry, we by no means claim a universal scaling law for all sizes of graphene ribbons in this work. It is the device size relative to the physical dimensions of the disorder that dictates the impact of disorder on carrier transport. Figure 1c can be viewed as a survey of $\sigma_{min}$ as a function of $W$ (with a constant $L$). As long as the transport is dominated by disorder (indeed for $L \geq 1$ μm), the key features of a peak and saturation minimum conductivity should be unchanged. We present the modeling results for the impact of edge disorder on devices with a different length ($L = 100$ nm, Figure S1), where the peak and saturation features remain but the peak height changes. A more complete length scaling for $\sigma_{peak}$ is shown in Fig 2b.

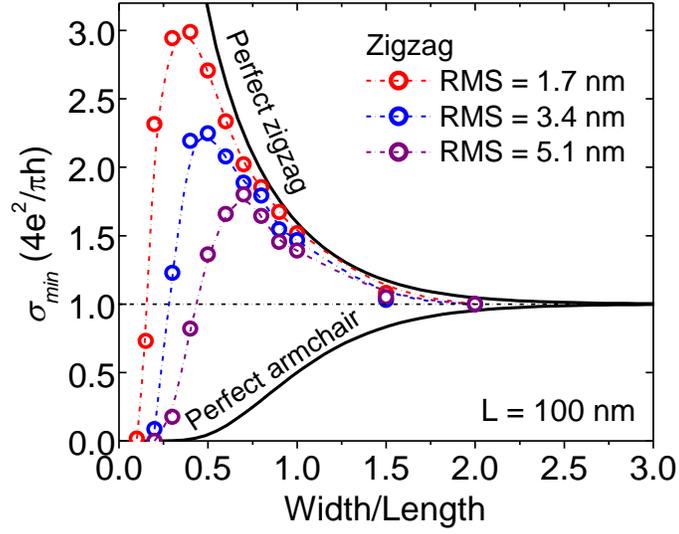

**Figure S1.** $\sigma_{min}$ vs. $W/L$ for edge-disordered graphene ribbons of $L = 100$ nm. The symbols are obtained from quantum simulations of zigzag ribbons with different edge roughness.

**II. Transmission Line Measurement (TLM) for contact resistance and transfer length**

Two-terminal graphene devices with $W \sim 800$ nm and $L = 500\text{-}2000$ nm are used to extract contact resistance and transfer length. The bottom gate is biased at the Dirac point to achieve maximum resistance in the graphene channels. The measurements are taken under vacuum ($1 \times 10^{-7}$ torr) at room temperature. $R_C \approx 500$ $\Omega \cdot \mu m$ and $L_T \approx 150$ nm are extracted from the TLM results (Figure S2), which are consistent with previous reports[4,5]. The electrical length of the channel is defined as $L \equiv L_{SEM} + 2L_T$. Considering the size of the source and drain contacts, the contact resistance is less than 5% of the total resistance for the off-state and has been subtracted from the total resistance for calculating $\sigma_{min}$.

**III. Temperature dependence for $\sigma_{min}$ in graphene ribbons with edge disorder and bulk disorder**

Conductance at finite temperature in the Landauer's formalism is $G = \int \partial f / \partial E \cdot T(E)\, dE$, where $T(E)$ is transmission. The temperature effect is incorporated in Fermi distribution f. In the zero disorder limit, the transmission of edge states follows $T(E) = C$, where C is proportion to the number of available edge states and is independent of energy E. Therefore, the edge states are temperature independent for disorder-free ribbons, since $G = C \int \partial f /\partial E \, dE = C$. The bulk states transmission follows $T(E) \propto E$. The temperature dependence of bulk states is discussed as follows.

For edge-disordered graphene ribbons, the temperature dependence is different for ribbons of different $W/L$ ratios. For ribbons with $W/L \sim 0.5$, $\sigma_{min}$ reaches the peak (Figure 1c) and the conduction is mainly through the edge states. The average device length is $L \sim 1.75$ μm, which gives us a width of $W \sim 0.875$ μm. This translates to an effective transport gap for the edge states of $\varepsilon_{loc} \approx 0.1$ meV[6]. To first order, we assume that the $T(E) = C$ if $|E| > \varepsilon_{loc}$, otherwise 0. Then one finds that $G$ changes from C to ~0.86C when temperature changes from 300 K to 4 K, which is a weak dependence.

For ribbons with $W/L \rightarrow 0$ or $W/L > 1$, the bulk states dominate the transport. In this regime, simple calculation indicates that the local potential fluctuation induced by bulk disorder is much larger than $k_B T$ at 300 K, as explained by the following. One could define a critical temperature $T^*$, above which the potential fluctuation induced by interface charge $n_s$ would be smaller than $k_B T$. To the first order, $T^* = \hbar v_F / k_B \cdot (\pi n_s)^{1/2}$, which translate to $n_s \approx 0.5 \times 10^{11}$ cm$^{-2}$ for $T^* \approx 300$ K. A typical $n_s$ value for graphene on SiO$_2$[7] is $\sim 2.3 \times 10^{11}$ cm$^{-2}$, which results in negligible temperature dependence due to bulk disorder. Therefore, neither edge disorder nor bulk disorder should cause appreciable temperature dependence in $\sigma_{min}$ in our parameter space window, as confirmed by our devices measured at T = 4-300 K (Figure 1b).

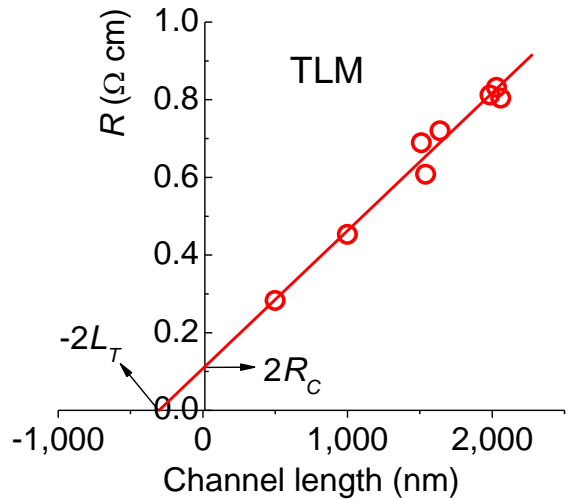

**Figure S2.** Measurement for graphene contact resistance and transfer length.

**IV. Edge disorder and bulk disorder models**

The tight-binding Hamiltonian describing the graphene sheet is given by:

$$H = \sum_i v_i \, a_i^\dagger a_i + \sum_{ij} t_{ij} \, a_i^\dagger a_j \quad (1)$$

where $a_i^\dagger/a_i$ are the creation/destruction operators at each atomic site i. $v_i$ and $t_{ij}$ are the on-site potential energy and hopping energy, respectively[8,9]. The quantum transport model is described elsewhere[10].

Edge roughness was recently characterized by High-Resolution Transmission Electron Microscopy, where an RMS ~ 3 nm was observed in exfoliated graphene ribbons[11]. In the presence of edge disorder, the coupling energy, $t_{ij}$ (unit eV), is modified as follows:

$$t(x,y) = \begin{cases} 3, & I_b(x) < y < W + I_t(x) \\ 0, & \text{otherwise} \end{cases} \quad (2)$$

$I_b$ and $I_t$ are one-dimensional roughness profiles describing the line edge roughness for the bottom and top edges, respectively[12]. In our edge construction, we have allowed the possibility of the so-called Klein defect[13,14], namely dangling carbon bonds. The edge roughness morphology is assumed to follow an exponential power spectrum. Two parameters, *RMS* and *AL*, are used to characterize the edge disorder. Figure S3a shows the power spectrum of the edge roughness used in our simulation.

Bulk disorder in graphene is believed to be a result of the spatial charge fluctuation at the interface between graphene and the underlying $SiO_2$ substrate[7]. The charge fluctuation is found to follow a Gaussian distribution. Standard deviation of the spatial charge density distribution and the effective puddle radius are the key parameters in the model to characterize the bulk disorder. We first generate a set of points uniformly distributed over the graphene sheet, i.e., $\vec{O}_i = (x_i, y_i)$. The electron-hole density is then given by:

$$n_0(\vec{r}) = \sum_{i=1}^{N} c_i \exp\left(-\frac{|\vec{r}-\vec{O}_i|^2}{2\mu_R^2}\right) \quad (3)$$

where $c_i$ is a random number following a Gaussian distribution from the Box-Muller scheme[15]. This translates to a potential of $|v(\vec{r})| = \hbar v_F / e\sqrt{\pi |n_0(\vec{r})|}$, with its sign following that of $n_0(\vec{r})$. Figure S3b and S3c display the statistical charge density distribution and the 2D potential landscape in a graphene ribbon with bulk disorder, respectively. Each puddle is assumed to have a Gaussian profile and interspersed randomly across the graphene ribbon.

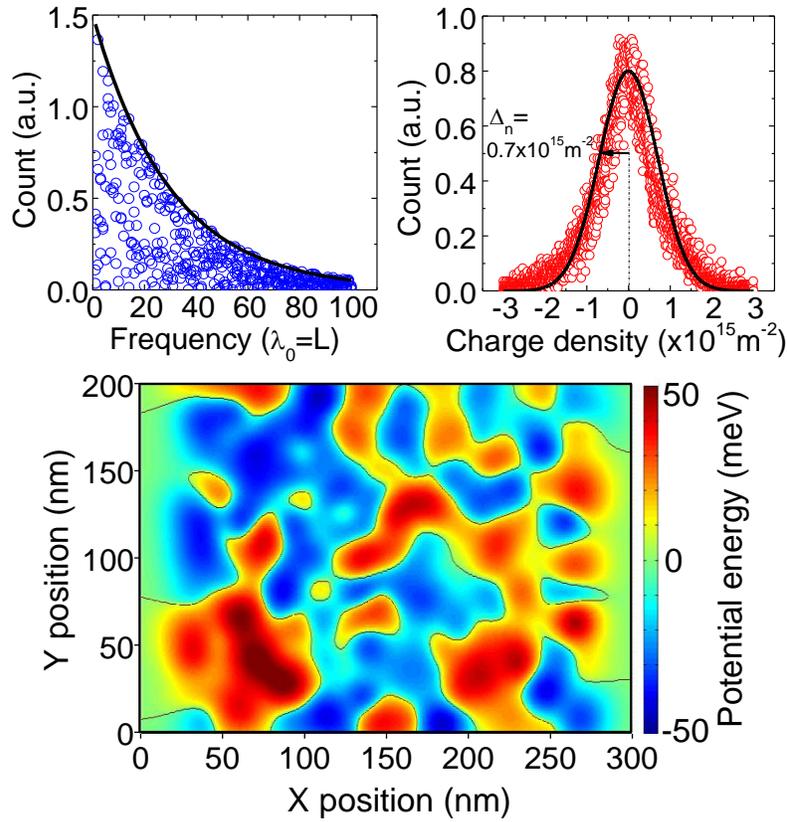

**Figure S3.** (a) The power spectrum of edge disorder profile for five statistical samples. (b) Spatial charge distribution across the graphene ribbon for five statistical samples. (c) Illustration of the potential landscape of a graphene ribbon with electron-hole puddles.

## V. Discussion for $\sigma_{min}$ when $W/L$ approaches zero

This work focuses graphene ribbons with $W \geq 100$ nm, where the energy gap due to size quantization is negligible. Therefore, the $\sigma_{min}$ value at $W/L = 0$ is beyond the scope of this study. A non-zero $\sigma_{min}$ as $W/L$ approaches zero is observed (Figure 1c), which is attributed to conduction through the bulk states in the presence of bulk disorder. In experimental graphene ribbons with $W/L \to 0$, edge disorder completely suppresses the propagation of the edge states, making bulk states the only ones responsible for current conduction. In our modeling work (Figure 2a, 3a, S1), we accordingly ignore the data for $W/L < 0.1$, where the transport is dominated by different mechanisms (e.g. bandgap due to size quantization).

## VI. Calculating the effective conductivity from the effective medium theory

$\sigma_1$ and $\sigma_2$ are defined to be the regions with high and low conductivity in the channel in the off-state as shown in Figure 3a. It is reasonable to assume equal areas of the $\sigma_1$ and $\sigma_2$ regions. From the classical effective medium theory[16], we obtain expressions for the effective classical conductivity as follows:

$$\sigma_{eff} \approx \left( \frac{1}{\sigma_1} + \frac{1}{\sigma_2} \right)^{-1} + \eta W^{\gamma}, \text{ for } W/L < 1 \qquad (4)$$

$$\sigma_{eff} \approx \sqrt{\sigma_1 \sigma_2}, \text{ for } W \gg \mu_R \text{ and } L \gg \mu_R \qquad (5)$$

$\eta$ and $\gamma$ are parameters describing the width scaling property of the system, which can be extracted from a binary resistor network. The value of $\sigma_2$ is determined by the graphene pn-interface and the puddle geometry. Assuming $\sigma_1 = 30$ a.u. and $\sigma_1/\sigma_2 = 0.01$, we estimate $\eta \approx 2.7$ and $\gamma \approx 0.16$. Figure 3c displays $\sigma_{eff}$ vs. $W/L$ according to the equations above. It captures the main features of the tight-binding simulation, namely an enhanced background conductivity that saturates at large $W/L$. The amount of $\sigma_{eff}$ enhancement from $4e^2/\pi h$ is governed by variables related to the bulk disorder, namely $\sigma_1$ and $\sigma_2$.

**VII. Comparing $\sigma_{peak}$ and $\sigma_{sat}$ between experiments and NEGF simulations**

Both edge disorder and bulk disorder should be considered when comparing the modeling results with the experimental observations. Bulk disorder enhances the saturation conductivity, with $\sigma_{sat}$ (modeling, $L$ = 1 μm) ≈ 4.3 ($4e^2/\pi h$) as computed using NEGF formalism, in good agreement with $\sigma_{sat}$ (experiment) ≈ 3.8 ($4e^2/\pi h$), as shown in Figure 1c.

When considering the peak height from the simulation, the enhancement from bulk disorder should be included since it impacts ribbons of all geometries. Therefore, $\sigma_{peak, overall}$ = $\sigma_{peak}$ (modeling, $L$ = 1 μm) + $\sigma_{sat}$ (modeling, $L$ = 1 μm) – ($4e^2/\pi h$) = 7.5 ($4e^2/\pi h$), where $\sigma_{peak}$ (modeling, $L$ = 1 μm) = 4.2 ($4e^2/\pi h$) according to Figure 2b. ($4e^2/\pi h$) is subtracted since it has been counted twice in $\sigma_{peak}$ and $\sigma_{sat}$. $\sigma_{peak, overall}$ is in quantitative agreement with the experimental observation of $\sigma_{sat}$ (experiment) = 7.8 ($4e^2/\pi h$), as shown in Figure 1c.

REFERENCES:

1. Miao, F.; Wijeratne, S.; Zhang, Y.; Coskun, U. C.; Bao, W.; Lau, C. N. *Science* **2007**, 317, 1530.

2. Tworzydlo, J.; Trauzettel, B.; Titov, M.; Rycerz, A.; Beenakker, C. W. J. *Physical Review Letters* **2006**, 96, 246802.

3. Danneau, R.; Wu, F.; Craciun, M. F.; Russo, S.; Tomi, M. Y.; Salmilehto, J., Morpurgo, A. F. *Physical Review Letters* **2008**, 100, 196802.

4. Chen, Z.; Appenzeller, J. *IEEE International Electron Devices Meeting Technical Digest*, **2008**, 509.

5. Nagashio, K.; Nishimura, T.; Kita, K.; Toriumi, A. *arXiv:1008.1826*.

6. Mucciolo, E. R.; Neto, A. H. C.; Lewenkopf, C. H. *Physical Review B* **2009**, 79, 075407.

7. Martin, J.; Akerman, N.; Ulbricht, G.; Lohmann, T.; Smet, J. H.; Von Klitzing, K.; Yacoby, A. *Nature Physics* **2008**, 4, 144.

8. Wallace, P. R. *Physical Review* **1947**, 71, 622.

9. Saito, R.; Dresselhaus, G.; Dresselhaus, M. S. *Physical Properties of Carbon Nanotubes* **1998** Imperial College Press.

10. Low, T.; Appenzeller, J. *Physical Review B* **2009**, 80, 155406.

11. Gupta, A. K.; Russin, T. J.; Gutierrez, H. R.; Eklund, P. C. *Acs Nano* **2009**, 3, 45.

12. Low, T. *Physical Review B* **2009**, 80, 205423.

13. Klein, D. J.; Bytautas, L. *Journal of Physical Chemistry A* **1999**, 103, 5196.

14. Kobayashi, Y.; Fukui, K.; Enoki, T.; Kusakabe, K.; Kaburagi, Y. *Physical Review B* **2005**, 71, 193406.

15. Box, G. E. P.; Muller, M. E. *Annals of Mathematical Statistics* **1958**, 29, 610.

16. Kirkpatrick, S. *Reviews of Modern Physics* **1973**, 45, 574.